\documentclass[journal]{IEEEtran}
\usepackage{subfigure}
\usepackage{graphicx}
\usepackage{array}
\usepackage{amsmath}
\usepackage{multirow}
\usepackage{ragged2e}
\usepackage{subfig}
\usepackage{color}
\usepackage{float}
\usepackage[ruled,linesnumbered]{algorithm2e}

\ifCLASSOPTIONcompsoc
  \usepackage[nocompress]{cite}
\else
  \usepackage{cite}
\fi

\begin{document}

\title{Approximate \textit{k}-NN Graph Construction:\\ a Generic Online Approach}

\author{{Wan-Lei Zhao*, Hui Wang, Chong-Wah Ngo}
\thanks{Wan-Lei Zhao and Hui Wang are with Fujian Key Laboratory of Sensing and Computing for
Smart City, Xiamen University, China. Wan-Lei Zhao is the corresponding author, email: wlzhao@xmu.edu.cn.}
\thanks{Chong-Wah Ngo is with Computer Science Department, City University of Hong Kong..}}

\IEEEtitleabstractindextext{
\begin{justify}
\begin{abstract}
Nearest neighbor search and \textit{k}-nearest neighbor graph construction are two fundamental issues arise from many disciplines such as multimedia information retrieval, data-mining and machine learning. They become more and more imminent given the big data emerge in various fields in recent years. In this paper, a simple but effective solution both for approximate \textit{k}-nearest neighbor search and approximate  \textit{k}-nearest neighbor graph construction is presented. These two issues are addressed jointly in our solution. On the one hand, the approximate \textit{k}-nearest neighbor graph construction is treated as a search task. Each sample along with its \textit{k}-nearest neighbors are joined into the \textit{k}-nearest neighbor graph by performing the nearest neighbor search sequentially on the graph under construction. On the other hand, the built \textit{k}-nearest neighbor graph is used to support \textit{k}-nearest neighbor search. Since the graph is built online, the dynamic update on the graph, which is not possible from most of the existing solutions, is supported. This solution is feasible for various distance measures. Its effectiveness both as \textit{k}-nearest neighbor construction and \textit{k}-nearest neighbor search approaches is verified across different types of data in different scales, various dimensions and under different metrics.
\end{abstract}
\end{justify}

}

\maketitle

\IEEEdisplaynontitleabstractindextext

\IEEEpeerreviewmaketitle

\section{Introduction}\label{sec:introduction}
Given a dataset $S$, \textit{k}-NN graph refers to the structure that keeps top-\textit{k} nearest neighbors for each sample in the dataset. It is the key data structure in the manifold learning~\cite{isomap, lle}, computer vision, machine learning and multimedia  information retrieval, etc~\cite{weidong}. Due to the fundamental role it plays, it has been studied for several decades.
Basically, given a metric, the construction of \textit{k}-NN graph is to find the top-\textit{k} nearest neighbors for each data sample. When it is built in brute-force way, the time complexity is $O(d{\cdot}n^2)$, where  $d$ is the dimension and $n$ is the size of dataset. Due to the prevalence of big data issues in various contexts, both $d$ and $n$ could be very large.

Despite numerous progress has been made in recent years, the major issues latent in approximate \textit{k}-NN graph construction still remain challenging. First of all, many existing approaches perform well only on low-dimensional data. The scale of data they are assumed to cope with is usually less than one million. Moreover, most of approaches are designed under specific metric \textit{i.e.}, $\textit{l}_2$-norm. Only recent few works~\cite{paredes:06, weidong, cvpr16:ben} aim to address this issue in the generic metric spaces. Thanks to the introduction of NN-Descent in~\cite{weidong}, the construction time complexity has been reduced from $O(n^{1.94})$~\cite{paredes:06} to $O(n^{1.14})$ for data with low dimensionality (\textit{e.g.}, \textit{5})~\cite{weidong}.

Besides the major issues aforementioned, many existing approaches still face another potential problem. Namely, most of the approaches are designed to build approximate \textit{k}-NN graph for a fixed dataset. In practice, it is not unusual that the dataset changes from time to time. This is particularly the case for large-scale multimedia search tasks. For instance, the photos and videos in Flickr grow on a daily basis. Given \textit{k}-NN graph is adopted to support the content-based search and browse, it should allow the dynamic update on the \textit{k}-NN graph. Such that the newly uploaded images and videos could be retrieved and browsed over as soon as they are put into the repository. Another typical scenario happens in e-shopping website. In the website, new clothes are put on sale and old-fashioned products should be withdrawn from sale each day.

In these scenarios, one would expect the \textit{k}-NN graph that works behind should be updated dynamically. Unfortunately, for most of the existing approaches~\cite{weidong, efanna16,jmlr09, msraKNN}, the dataset is assumed to be fixed. Any update on the dataset invokes a complete reconstruction on the \textit{k}-NN graph. As a consequence, the aggregated costs are very high even the dataset is in small-scale. It is more convenient if it is allowed to simply insert/remove the samples into/from the existing \textit{k}-NN graph. Nevertheless, it is complicated to update the \textit{k}-NN graph with the support of conventional indexing structure such as locality sensitive hashing (LSH)~\cite{lsh04}, R-Tree~\cite{rtree84} or k-d tree~\cite{kdtree75}.

Another problem that is closely related to approximate \textit{k}-NN graph construction is nearest neighbor search (NN search), which also arises from a wide range of applications. The nearest neighbor search problem is defined as follows. Given a query vector ($q \in R^d$), and \textit{n} candidates in $S$ that are under the same dimensionality. It is required to return sample(s) that are the closest to the query according to a given metric $m(\cdot, \cdot)$.

In this paper, a generic approximate \textit{k}-NN graph construction approach is presented. The issues of \textit{k}-NN graph construction and NN search are addressed under a unified framework. The graph construction problem is treated as a \textit{k}-NN search task. The approximate \textit{k}-NN graph is incrementally built by invoking each sample to query against the graph under construction. After one round of \textit{k}-NN search, the query sample is joined into the graph along with the discovered top-\textit{k} nearest neighbors. The \textit{k}-NN lists of samples (already in the graph) that are visited during the search are accordingly updated. The NN search basically follows the hill-climbing strategy~\cite{icai11:kiana}. In order to achieve high performance in terms of both efficiency and quality, two major innovations are proposed.

\begin{itemize}
	\item {Restricted recursive neighborhood propagation (RRNP) is proposed to introduce the newly coming sample to its most likely neighbors, which enhances the quality of approximate \textit{k}-NN graph considerably.}
	\item {In order to boost the search performance, a lazy graph diversification (LGD) scheme is proposed. It helps to avoid unnecessary distance computations during the hill-climbing while involving no additional computations.}
\end{itemize}

The advantages of this approach are several folds. Firstly, the online construction avoids repetitive distance computations that most of the current approximate \textit{k}-NN graph construction approaches suffer from. This makes it more efficient than the classic NN-Descent algorithm~\cite{weidong}. Secondly, the online graph construction is particularly suitable for the scenario that dataset is dynamically changing. Moreover, two closely related issues, namely \textit{k}-NN graph construction and NN search, have been jointly addressed in our solution. Compared to the state-of-the-art NN search approaches~\cite{dpg:wenli,pami18:yury,efanna16}, it shows similar or even slightly higher search efficiency while maintaining an online \textit{k}-NN graph. The advantage is that it allows the user to browse over hyper-links between closely related contents (\textit{i.e.}. \textit{k} nearest neighbors). Furthermore, our approach has no specification on the distance measure, it is therefore a generic solution, which is confirmed in our experiments.

The remainder of this paper is organized as follows. In Section~\ref{sec:rela}, a brief review about the research works on approximate \textit{k}-NN graph construction and approximate \textit{k}-NN search is presented. Section~\ref{sec:nsw} presents an NN search algorithm upon which the approximate \textit{k}-NN graph construction approach is built. Section~\ref{sec:knn} presents an online approximate \textit{k}-NN graph construction approach and the enhancement schemes. A more efficient NN search approach based on the online graph is presented in Section~\ref{sec:nns}. The experimental studies about the effectiveness of proposed \textit{k}-NN graph construction and NN search are presented in Section~\ref{sec:exp}. Section~\ref{sec:end} concludes the paper.

\section{Related Works}
\label{sec:rela}
\subsection{\textit{k}-NN Search}
The early study about the \textit{k}-NN search issue could be traced back to \textit{1970s} when the need of NN search on the file system arises. In those days, the data to be processed are in very low dimension, typically 1D. This problem is well-addressed by B-Tree~\cite{btree:commer79} and its variant $B^+$-tree~\cite{btree:commer79}, based on which the NN search time complexity could be as low as $O(log(n))$. B-tree is not naturally extensible to more than 1D case. More sophisticated indexing structures were designed to handle NN search in multi-dimensional data. Representative structures are k-d-tree~\cite{kdtree75}, R-tree~\cite{rtree84} and R*-tree~\cite{BKSB90}. For k-d tree, pivot vector is selected each time to split the dataset evenly into two. By applying this bisecting repeatedly, the hyper-space is partitioned into embedded hierarchical subspaces. The NN search is performed by traversing over one or several branches to probe the nearest neighbors. Unlike B-tree in 1D case, the partition scheme does not exclude the possibility that nearest neighbor resides outside of these candidate subspaces. Therefore, extensive probing over the large number of branches in the tree becomes inevitable. For this reason, NN search with k-d tree and the like could be very slow. Recent indexing structure FLANN~\cite{pami14:flann,cvpr08:annoy} partitions the space with hierarchical \textit{k}-means and multiple k-d trees. Although efficient, sub-optimal results are observed.

For all the aforementioned tree partitioning methods, another major disadvantage lies in their heavy demand in memory. On the one hand, in order to support fast comparison, all the candidate vectors are loaded into the memory. On the other hand, the tree nodes that are used for indexing also take up considerable amount of extra memory. Overall, the memory consumption is usually several times bigger than the size of reference set.

In order to reduce the memory complexity, quantization based approaches~\cite{ivfrvq10, JDS11,artem16,icml14:tzhang, sq14} are proposed to compress the reference vectors~\cite{tit06:gray06,ivfrvq10}. For all the quantization based methods, they share two things in common. Firstly, the candidate vectors are all compressed via vector (or sub-vector) quantization. Secondly, NN search is conducted between the query and the compressed candidate vectors. The distance between query and candidates is approximated by the distance between query and vocabulary words that are used for quantization. Due to the heavy compression on the reference vectors, high search quality is hardly achievable. Furthermore, these types of approaches are only suitable for metric spaces of $\textit{l}_p$-norm.

Apart from above approaches, several attempts have been made to apply LSH~\cite{lsh04, mlsh07} on NN search. In general, there are two steps involved in the search stage. Namely, \textit{step 1} collects the candidates that share the same or similar hash keys as the query. \textit{Step 2} performs exhaustive comparison between the query and all these selected candidates to find out the nearest neighbor. Similar as FLANN, computational cost remains high if one expects high search quality. Additionally, the design of hash functions that are feasible for various metrics is non-trivial.

Recently, the graph-based approaches have been extensively explored~\cite{icai11:kiana,mm12:jdwang,infosys13:yury, cvpr16:ben, efanna16,vldb19:fu}. Although they are different for each other in details, all of them are built upon a hill-climbing procedure. The search~\cite{icai11:kiana} starts from a group of random seeds (random locations in the vector space). It traverses iteratively over an approximate \textit{k}-NN graph or a relative \textit{k}-NN graph (built in advance) by best-first search. Guided by the NN graph, the search procedure ascends closer to the true nearest neighbor in each round until no better candidates could be found. Approaches in~\cite{icai11:kiana,pami18:yury,infosys13:yury,efanna16,vldb19:fu} follow similar search procedure. The major difference between them lies in the graph used to support the search. According to recent study~\cite{dpg:wenli}, these graph-based approaches demonstrate superior performance over other types of approaches across variety types of data.

\subsection{Approximate \textit{k}-NN Graph Construction}
The approaches for approximate \textit{k}-NN graph construction can be roughly grouped into two categories. Approaches such as~\cite{msraKNN,pkdd13:zhang} basically follow the divide-and-conquer strategy. On the first step, samples are partitioned into a number of small subsets. Exhaustive comparisons are carried out within each subset. The closeness relations (\textit{viz.}, edges in the \textit{k}-NN graph) between any two samples in one subset are established. In the second step, these closeness relations are collected to build the \textit{k}-NN graph. To enhance the performance, the first step is repeated for several times with different partitions. The produced closeness relations are used to update the \textit{k}-NN graph. Since it is hard to design partition scheme that is feasible for various generic spaces, they are generally only effective in $\textit{l}_2$-space. Another category of approximate \textit{k}-NN graph construction, typically NN-Descent~\cite{weidong} avoids such disadvantage. The graph construction in NN-Descent starts from a random \textit{k}-NN graph. Based on the principle ``neighbor's neighbor is likely to be the neighbor'', the \textit{k}-NN graph evolves by invoking comparison between samples in each sample's neighborhood. Better closeness relations that are produced in the comparison are used to update the neighborhood of one sample. This approach turns out to be generic and efficient. Essentially, it can be viewed as performing hill-climbing batchfully~\cite{weidong}. Recently, the mixture scheme derived from the  above approaches is also seen in the literature~\cite{efanna16}.

It is worth noting that approaches proposed in~\cite{pami18:yury,cvpr16:ben, dpg:wenli} are not approximate \textit{k}-NN graph construction algorithms. The graphs are built primarily for \textit{k}-nearest neighbor search. In these approaches, the samples which should be in the \textit{k}-NN list of one sample are deliberately omitted for comparison efficiency. While the links to the remote neighbors are maintained~\cite{pami18:yury}. As a consequence, graphs constructed by these approaches are not \textit{k}-NN graph in the real sense. Such kind of graphs are hardly supportive for tasks beyond \textit{k}-NN search. Navigable small world graph (NSW)~\cite{infosys13:yury} is primarily designed to support fast NN search, it could be viewed as an online graph construction approach. However, it is not suitable for approximate \textit{k}-NN graph construction for its poor search strategy, which leads to low graph quality.

In most of the approaches aforementioned, one potential issue is that the construction procedure has to keep records on the comparisons that have been made between sample pairs to avoid repetitive comparison. However, its space complexity could be as high as $O(n^2)$. Otherwise the repetitive comparisons are inevitable even by adopting  specific sampling schemes~\cite{weidong}. In this paper, the approximate \textit{k}-NN graph construction and \textit{k}-NN search are addressed jointly. The approximate \textit{k}-NN graph construction is undertaken by invoking each sample as a query to query against the approximate \textit{k}-NN graph that is under construction. Since the query is new to the graph under construction each time, two samples in the dataset are compared at most once. The repetitive comparisons are avoided.

\section{NN Search on the \textit{k}-NN Graph}
\label{sec:nsw}
Before we introduce our graph construction approach, the NN search, on which the construction approach is based, is presented. Our search procedure is largely similar as the procedure proposed in HNSW~\cite{pami18:yury}. Whereas unlike HNSW, our search procedure is undertaken on a flat \textit{k}-NN graph instead of a hierarchical relative neighborhood graph. Compared to the single-layer search in HNSW, a few modifications are further introduced. In the flat \textit{k}-NN graph, both the \textit{k} nearest neighbors of one sample and its reverse neighbors are kept. In the following, the structure of this graph is discussed before we present the search procedure.
\begin{figure}
	\begin{center}
	\subfigure[a graph with \textit{4} vertices]
    {\includegraphics[width=0.36\linewidth]{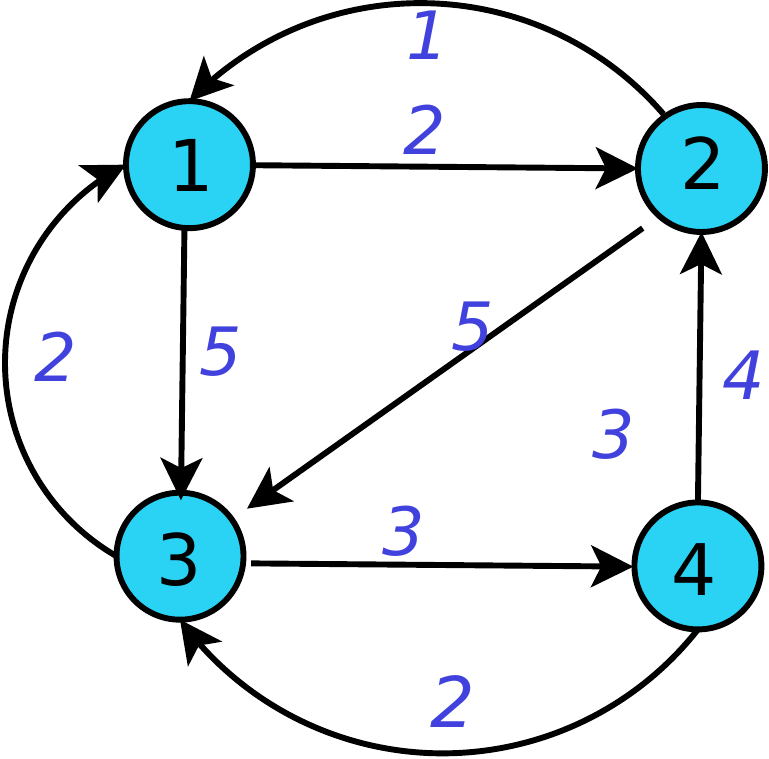}} \\
	\subfigure[\textit{2}-NN graph $G$ for (a)]
	{\includegraphics[width=0.36\linewidth]{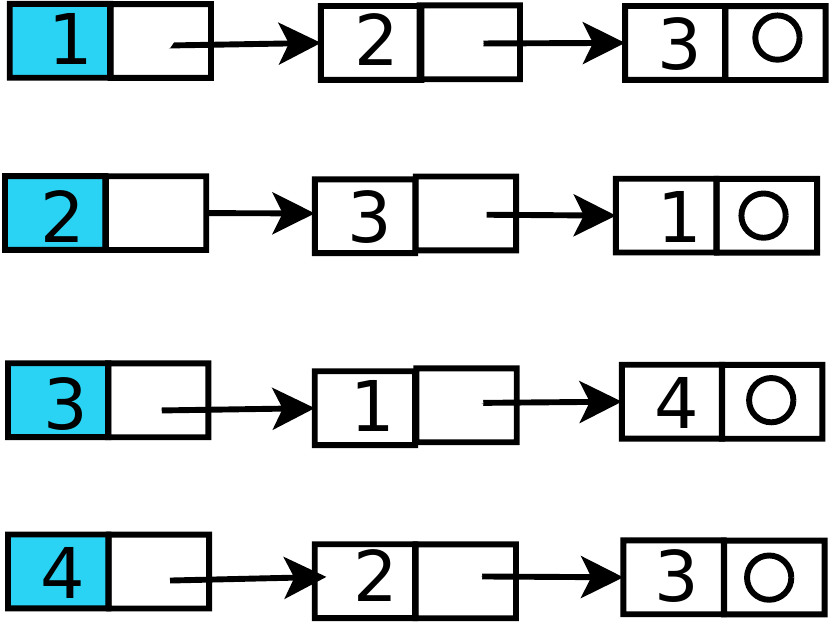}}
    \hspace{0.30in}
	\subfigure[reverse \textit{2}-NN graph $\overline{G}$ for (a)]
    {\includegraphics[width=0.49\linewidth]{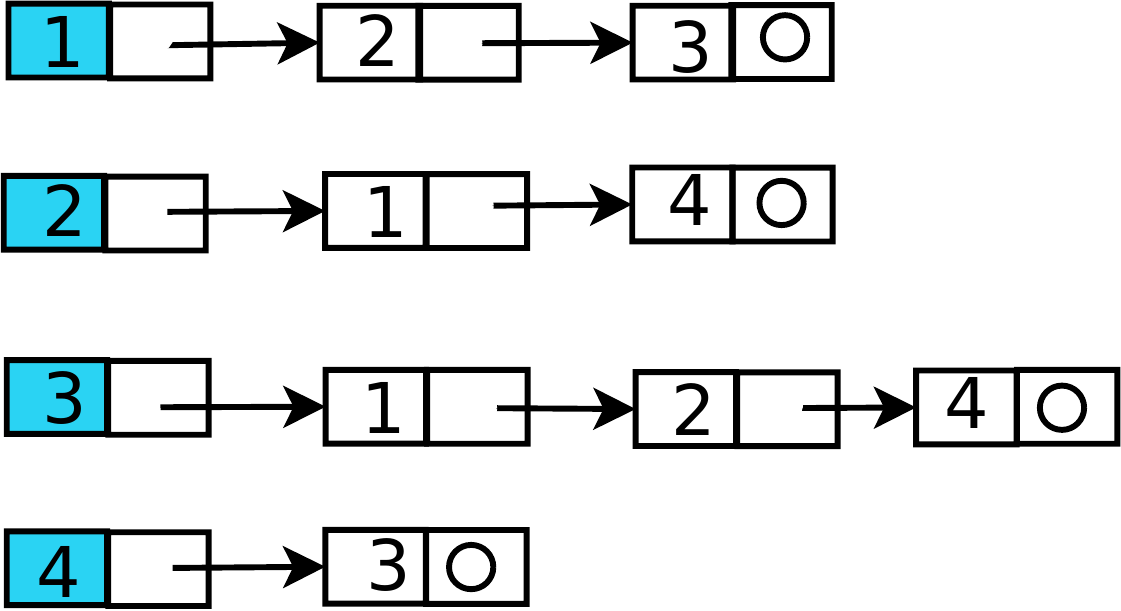}}
	\caption{An illustration of \textit{k}-NN graph $G$ and its reverse \textit{k}-NN graph $\overline{G}$. In the illustration, $k$ is set to \textit{2}.}
	\label{fig:knn1}
\end{center}
\end{figure}

Given \textit{k}-NN graph $G$, $G[i]$ returns \textit{k}-NN list of sample $i$. Accordingly, $\overline{G}$ is the reverse \textit{k}-NN graph of $G$, which is nothing more than a re-organization of graph $G$. $\overline{G}[i]$ keeps ID of samples that sample $i$ appears in their $k$-NN lists. Noticed that the size of $\overline{G}[i]$ is not necessarily $k$ and there would be overlappings between $\overline{G}[i]$ and $G[i]$. An illustration about graphs $G$ and $\overline{G}$ are  seen in Fig.~\ref{fig:knn1}. In our implementation, the union of $G[i]$ and $\overline{G}[i]$ are kept in a dynamic array (instead of linked list). The first \textit{k} elements are the \textit{k} nearest neighbors (namely $G[i]$). The remaining elements are the reverse neighbors of sample \textit{i} that are outside the coverage of \textit{k} nearest neighbors. For presentation clarity, these samples in the neighborhood are still referred as $G[i]$ and $\overline{G}[i]$. With the support of \textit{k}-NN graph $G$ and its reverse graph $\overline{G}$, the NN search algorithm is presented in Alg.~\ref{alg:nswsearch}.

\begin{algorithm}
  \KwData{\textit{q}: query; $G$: \textit{k}-NN Graph; $\overline{G}$: reverse \textit{k}-NN Graph; $S_{n{\times}d}$: reference set; $k$: num. of NN; $p$: num. of seeds}
  \KwResult{\textit{Q}: \textit{k}-NN list of \textit{q};
			\textit{V}: Samples visited during the search;
			\textit{D}: Distances between sample \textit{q} and the samples visited during the search}
	$Q \leftarrow \emptyset$; D[1$\cdots$n] $\leftarrow \infty$\;
	Flag[1$\cdots$n]$\leftarrow 0$; R[1{$\cdots$}p] $\leftarrow$ \textit{p} random seeds\;
    \For {each $r \in$ \textit{R}}{
		Flag[$r$] $\leftarrow$ 1; D[r] $\leftarrow m(r, q)$\;
		\textit{f} $\leftarrow$ get furthest sample to \textit{q} from \textit{Q} \;
		\If {$m(r, q) < m(f, q)$ or $|Q| < k$} {
			$Q \leftarrow Q \cup r$\;
			$R \leftarrow R \cup r$\;
		}
	}
	\While {$ |R| > 0$} {
		\textit{r} $\leftarrow$ pop nearest sample to \textit{q} from \textit{R}\;
		\textit{f} $\leftarrow$ get furthest sample to \textit{q} from \textit{Q} \;
		\If{$m(r, q) > m(f, q)$} {
			break\;
		}
		\For{each $e \in G[r] \cup \overline{G}[r]$} {
			\If {Flag[$e$] == 0} {
				Flag[$e$] $\leftarrow$ 1; D[$e$] $\leftarrow m(e, q)$\;
				\textit{f} $\leftarrow$ get furthest sample to \textit{q} from \textit{Q}\;
				\If {$m(e, q) < m(f, q)$ or $|Q| < k$} {
					$Q \leftarrow Q \cup e$\;
					$R \leftarrow R \cup e$\;
				}
			}
		}
	}
	Collect the visited samples marked in \textit{Flag} to \textit{V}\;
 \caption{NN Search on Graph (NNSearch($\cdot$))}
 \label{alg:nswsearch}
\end{algorithm}

Alg.~\ref{alg:nswsearch} in general is a hill-climbing procedure~\cite{icai11:kiana} with multiple starting seeds. The query $q$ is firstly compared to \textit{p} seed samples. The compared samples are kept in two priority queues $Q$ and $R$. $Q$ basically maintains the top-\textit{k} nearest neighbors of the query $q$. The retrieved neighbors are sorted in ascending order according to their distance to the query\footnote{Without the loss of generality, it is assumed that the smaller of the distance the closer of two samples across the paper.}. As a closer sample is joined into Q, the rear sample in $Q$ will be swapped out. Unlike $Q$, the size of $R$ is not fixed. 
In each iteration (\textit{Line 13 -- 31}), the algorithm visits the neighborhood of the closest sample $r \in R$ to the query $q$. $Q$ and $R$ are updated accordingly with new close samples to the query. The iteration continues until $R$ is empty or $Q$ cannot be updated. In the comparison, the distance function $m(\cdot,\cdot)$ could be any metric defined on the input dataset. It is clear to see this is a generic search algorithm. The major difference between Alg.~\ref{alg:nswsearch} and single-layer NN search algorithm from HNSW~\cite{pami18:yury} are in two aspects. Firstly, it starts from multiple random seeds, which leads to more stable performance over HNSW. Moreover, the visited samples and the corresponding distances to the query are kept. These distances will be used to assist the online \textit{k}-NN graph diversification later.

\section{Online Approximate \textit{k}-NN Graph Construction}
\label{sec:knn}

The prerequisites for NN search algorithm in Alg.~\ref{alg:nswsearch} are the \textit{k}-NN graph $G$ and its reverse \textit{k}-NN graph $\overline{G}$. In this section, we are going to show how an approximate \textit{k}-NN graph and its reverse graph are built based on NN search algorithm itself. Additionally, a strategy called \textit{restricted recursive neighborhood propagation} is proposed to enhance the quality of approximate \textit{k}-NN graph. Moreover, an online graph diversification scheme is proposed. Compared to the approaches in~\cite{dpg:wenli, cvpr16:ben}, it involves no additional distance computations while leading to more efficient NN search than Alg.~\ref{alg:nswsearch}.

\subsection{Approximate \textit{k}-NN Graph Construction by Search}
In Alg.~\ref{alg:nswsearch}, the search starts from several random locations of the reference set and moves along several trails. The search moves towards the closer neighborhood of a query in each iteration. In the ideal case, top-\textit{k} nearest neighbors will be discovered. On the one hand, the top-\textit{k} nearest neighbors of this query are known after the search. On the other hand, some samples in the reference set are introduced with a new neighbor (namely the query). As a result, the \textit{k}-NN graph could be augmented to include this query sample.

Motivated by this observation, the online \textit{k}-NN graph construction algorithm is conceived. First, an initial graph is built exhaustively from a small subset of $S$. The size of $S$ is fixed to \textit{64} in this paper. After that, each of the remaining samples is treated as query to query against the \textit{k}-NN graph following the flow of Alg.~\ref{alg:nswsearch}. The \textit{k}-NN list of a query sample is joined into the graph after each search. The \textit{k}-NN lists of samples which have been visited during the search are accordingly updated. The general procedure of the construction algorithm is summarized in Alg.~\ref{alg:knnbuild1}.

\begin{algorithm}
    \KwData{$S_{n{\times}d}$: dataset; $k$: num. of NN; $p$: num. of seeds}
    \KwResult{$G$: \textit{k}-NN Graph}
    \textit{Extract} a small subset $I$ from $S$\;
    \textit{Initialize} ${G}$ and $\overline{G}$ with $I$\;
    \For {each $q \in$ ${S-I}$}{
		$Q, V, D \leftarrow $NNSearch($q, G, \overline{G}, S_{n{\times}d}, k, p$)\;
		/*\textit{V} is the set of samples visited during the search*/\\
		/*\textit{D} keeps distances btw. \textit{q} and the visited samples*/\\
		\For {each $r \in$ \textit{V}} {
			InsertG($r, q, D[r], G$)\;
			InsertG($q, r, D[r], \overline{G}$)\;
		}
	 }
  \caption{Online Approximate \textit{k}-NN Graph Constr. (OLGraph)}
 \label{alg:knnbuild1}
\end{algorithm}

In Alg.~\ref{alg:knnbuild1}, the procedure of approximate \textit{k}-NN graph construction is basically a repetitive calling of the NN search algorithm and graph update function. Function \textit{InsertG}($\cdot$) is responsible for inserting an edge into \textit{k}-NN list of \textit{r} in graphs $G$ and $\overline{G}$. The major operations inside \textit{InsertG}($\cdot$) involve the update of \textit{k}-NN list and the reverse \textit{k}-NN list of \textit{r}. A sample in the rear of a \textit{k}-NN list is deleted if a closer sample is joined in. Distances of \textit{k}-NNs are kept to allow the list to be sorted all the time.

Although the size of input dataset is fixed in Alg.~\ref{alg:knnbuild1}, it is apparently feasible for an open set, for which new samples are allowed to join in from time to time. As will be revealed in the experiments (Section~\ref{sec:exp}), Alg.~\ref{alg:knnbuild1} already performs very well. In the following, two novel schemes are presented to further boost the performance in the graph construction and NN search respectively.

\subsection{Restricted Recursive Neighborhood Propagation}

\begin{figure}
\begin{center}
	\subfigure[depth=1]
    {\includegraphics[width=0.30\linewidth]{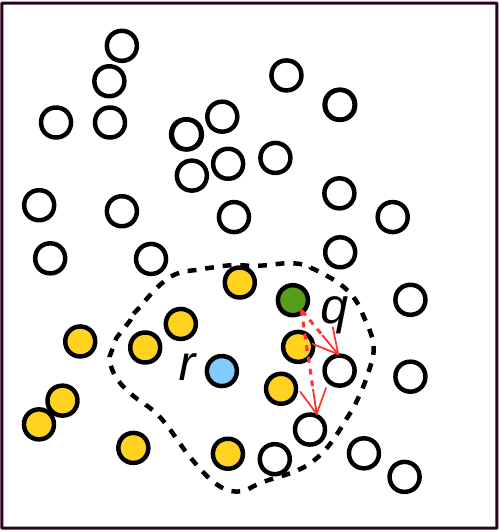}}
    \hspace{0.10in}
	\subfigure[depth=2]
	{\includegraphics[width=0.30\linewidth]{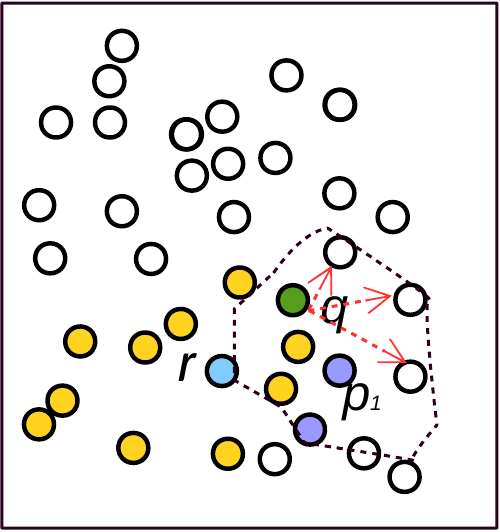}}\\
	\subfigure[depth=3]
    {\includegraphics[width=0.68\linewidth]{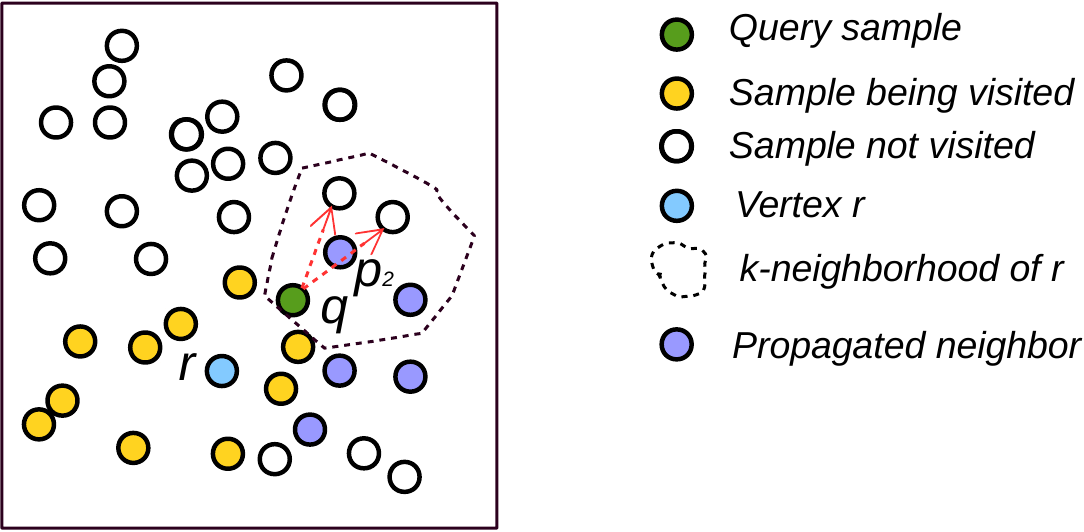}}
	\caption{The illustration of \textit{restricted recursive neighborhood propagation}. The propagation starts from unvisited samples in \textit{r}'s neighborhood and expands deeper to the neighbors of \textit{r}'s neighbor. In the figure, $p_1$ is \textit{r}'s neighbor, $p_2$ is $p_1$'s neighbor. Neither of them are encountered during the search.}
	\label{fig:rnp}
\end{center}
\end{figure}

In the last steps of Alg.~\ref{alg:knnbuild1}, the query sample $q$ is inserted to the neighborhood of each $r$ as long as $q$ is in their \textit{k}-NN range. Noticed that only a few \textit{r}s that are sufficiently close to $q$ will be considered. In the neighborhood of \textit{r}, it is possible that there are some samples that were not compared to $q$ during the hill-climbing search (shown as empty circles inside the dashed polygon of Fig.~\ref{fig:rnp}(a)). Based on the principle that ``neighbor's neighbor is likely to be the neighbor'', these unvisited samples are likely to be close neighbors of sample $q$. It is therefore reasonable to insert $q$ and these unvisited samples to the neighborhoods of each other. After this insertion, it is possible that $q$ is introduced to meet with more unvisited neighbors. As a result, such kind of insertion could be undertaken recursively until no new unvisited neighbors are encountered. In addition, such kind of propagation is restricted to the close neighborhood of a sample. For example, as shown in Fig.~\ref{fig:rnp}(b), $p_1$'s neighborhood is propagated only if $m(q,p_1)$ is smaller than the distance from $p_1$ to its \textit{k}-th neighbor. This operation is called as \textit{restricted recursive neighborhood propagation} (RRNP). We find this operation further improves the quality of \textit{k}-NN graph while only inducing minor computational overhead.

The procedure of RRNP is illustrated in Fig.~\ref{fig:rnp} and shown in Alg.~\ref{alg:knnbuild2} (\textit{Line 9 -- 24}). $W$ represents a working queue for the samples to be propagated. \textit{Push}($\cdot$) inserts a new element at the end of the queue and \textit{Pop}($\cdot$) removes and returns the next element in the queue. Firstly, a sample $r$ is pushed into $W$. After that, all neighbors of the sample(s) in $W$ that meet the conditions will be pushed into the queue, and the $k$-NN graph is updated accordingly. This operation will be repeated until the queue is empty, which means that there is no sample available for propagation.

Although RRNP is conceptually similar as neighborhood propagation proposed in~\cite{cvpr12:jingwang}, they are essentially different in two aspects. Firstly, there is no priority in terms of propagation in RRNP. All the unvisited samples in one neighborhood are compared with the query in random order. Furthermore, such kind of propagation is restricted to the close neighborhood (within the radius of a sample's \textit{k}-NN list). The neighborhood propagation proposed in~\cite{cvpr12:jingwang} is more like a mini version of NN-Descent.

\begin{algorithm}
    \KwData{$S_{n{\times}d}$: dataset; $k$: num. of NN; $p$: num. of seeds; $dp$: max. depth of propagation}
    \KwResult{$G$: \textit{k}-NN Graph with LGD information}
    \textit{Extract} a small subset $I$ from $S$\;
	\textit{Initialize} ${G}$ and $\overline{G}$ with $I$\;
    \For {each $q \in$ ${S-I}$}{
		$Q, V, D \leftarrow $NNSearch($q, G, \overline{G}, S_{n{\times}d}, k, p$)\;
		Flag[1$\cdots$n]$\leftarrow 0$\;
		\For {each $r \in$ \textit{V}} {
			InsertG($r, q, G$)\;
			InsertG($q, r, \overline{G}$)\;
			$W \leftarrow \emptyset$\;
			Push($r, depth(r), W$); //depth of $r$ is 0\\
			\While{$|W| > 0$} {
				$s \leftarrow $ Pop($W$)\;
				$f \leftarrow $ the \textit{k}-th neighbor of $s$\;
				\If{$depth(s) < dp$ \&\& $m(q, s) < m(s, f)$} {
					\For{each $e \in G[s] \cup \overline{G}[s]$} {
						\If {$e \notin V$ \&\& Flag[$e$] == 0} {
							Flag[$e$] $\leftarrow$ 1\;
							InsertG($e, q, G$)\;
							InsertG($q, e, \overline{G}$)\;
							Push($e, depth(s) + 1, W$)\;
						}
					}
				}
			}
			ApplyLGD($G[r], D$); //Apply \textit{LGD} rules\\
		}
	 }
  \caption{Online Approximate \textit{k}-NN Graph Constr. with RRNP and LGD (OLGraph$^+$)}
 \label{alg:knnbuild2}
\end{algorithm}

\subsection{Lazy Graph Diversification}
\begin{figure}
\begin{center}
    \includegraphics[width=0.42\linewidth]{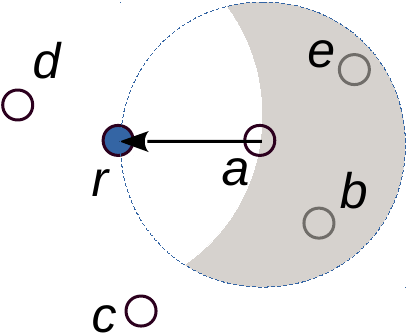}
    	\caption{An illustration of occlusion in 2D $l_2$-space happens in the neighborhood of sample $r$. Samples $a$, $b$, $c$, $d$ and $e$ are in the \textit{k}-NN list of $r$. $m(r, b)$ is greater than $m(r, a)$, while $m(a, b)$ is smaller than $m(r, b)$, we say that $b$ is occluded by $a$ in \textit{r}'s neighborhood, while sample $c$ and $d$ are not occluded by $a$. Actually all the samples located in the moon shape shadow are occluded by $a$. Notice that the region that is occluded by $a$ could be beyond this moon shape region.}
	\label{fig:shadow}
\end{center}
\end{figure}

In Alg.~\ref{alg:nswsearch}, when expanding sample $r$, all the samples in the neighborhood of $r$ will be compared to the query. According to recent studies~\cite{dpg:wenli,pami18:yury,cvpr16:ben}, when samples in the neighborhood of $r$ are very close to each other, it is no need to compare to all of them during the expansion. The expansion on these close samples most likely guides the climbing process to the same local region. The phenomenon that samples in the \textit{k}-NN list are closer to each other than they are to $r$ is called as ``occlusion''~\cite{cvpr16:ben}. An illustration of occlusion is shown in Fig.~\ref{fig:shadow}. In the illustration, samples $b$ and $e$ are occluded by sample $a$. It is easy to see one sample can only be occluded by samples which are closer to $r$ than that of it. According to~\cite{dpg:wenli,pami18:yury,cvpr16:ben}, the hill-climbing will be more efficient when samples like $b$ and $e$ are not considered when expanding $r$.

In order to know whether samples in a \textit{k}-NN list are occluded by each other, the pair-wise comparisons between samples in the \textit{k}-NN list are required~\cite{dpg:wenli,cvpr16:ben,pami18:yury}. This is unfeasible for an online construction procedure (\textit{i.e.}, Alg.~\ref{alg:knnbuild1}). First of all, \textit{k}-NN lists are dynamically changing, pair-wise distances cannot be simply computed and kept for use all the way. Secondly, it is too costly to update the pair-wise occlusion relations as long as a new sample is joined in. It would induce a complete comparison between this new sample and the rest. Moreover, unlike HNSW the occluded samples cannot be simply removed from a \textit{k}-NN list since our primary goal is to build an approximate \textit{k}-NN graph as precise possible.

In this paper, a novel scheme called \textit{lazy graph diversification} (LGD) is proposed to identify the occlusions between samples during the online graph construction. Firstly, an occlusion factor $\lambda$ is introduced as the attribute attached to each sample in a \textit{k}-NN list. $\lambda$s of all the samples in the list are initialized to \textit{0} when the \textit{k}-NN list of a new query is joined into the graph. Factor $\lambda$ will be updated when another new sample is joined into this \textit{k}-NN list at the later stages. 

Let's consider a new sample \textit{q} to be inserted into sample \textit{r}'s \textit{k}-NN list. In order to update $\lambda$s of \textit{r}'s neighbors, we should know the distances of all the neighbors to \textit{r} and the distances between \textit{q} and other neighbors in the list. The distances to \textit{r} are already known. While the distances between the query and the rest neighbors are unknown. Instead of performing a costly comparison between \textit{q} and the rest neighbors, we make use of distances that are recorded in variable \textit{D}. Namely, the distances between \textit{q} and all the visited samples during the NN search (Alg.~\ref{alg:nswsearch}). With the support of \textit{D}, occlusion factor $\lambda$ of all the samples in the \textit{k}-NN list is updated with following three rules.

\begin{itemize}
	\item {\textbf{Rule 1}: $\lambda$ is kept unchanged for samples ranked before \textit{q};}
	\item {\textbf{Rule 2}: $\lambda$ of sample $q$ is incremented by \textit{1} if a sample ranked before \textit{q} is closer to \textit{q} than \textit{q} to \textit{r};}
	\item {\textbf{Rule 3}: $\lambda$ of a sample ranked after \textit{q} is incremented by \textit{1} if its distance to \textit{q} is smaller than \textit{q} to \textit{r}.}
\end{itemize}

\begin{figure*}
\begin{center}
    \includegraphics[width=0.75\linewidth]{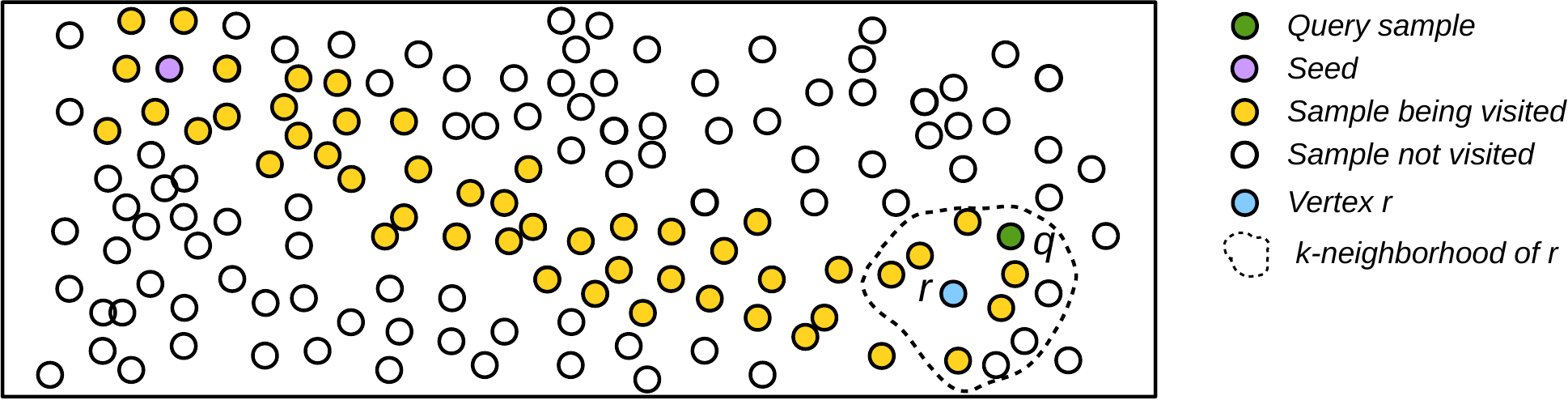}
	\caption{A trail of hill-climbing procedure in 2D $l_2$-space. The hill-climbing starts from a single seed and converges when it reaches to the neighborhood of the query. Query sample $q$ is to be inserted into \textit{k}-NN list of $r$. The occlusion relations between $q$ and the rest samples in $r$'s neighborhood have to be updated. The distances from samples in the list to $r$ are known. The distances from $q$ to visited samples in $r$'s neighborhood are also known. The distances from $q$ to not-being-visited are $\infty$. Based on the \textit{LGD rules}, occlusion factor $\lambda$s of samples in $r$'s neighborhood could be updated.}
	\label{fig:ldg}
\end{center}
\end{figure*}

The default distance of each sample to \textit{q} is set to $\infty$. The $\lambda$s of not-being-visited neighbors are not updated according to \textit{Rule 1} and \textit{Rule 3}. This is reasonable because the not-being-visited neighbors should be sufficiently far away from $q$, otherwise they are already being visited according to the principle ``a neighbor of a neighbor is also likely to be a neighbor''. Since we have all the possible distances (between $q$ and samples in the graph) only after the the hill-climbing converges, the operations of inserting $q$ into \textit{k}-NN list of $r$ and updating factor $\lambda$s in the list are postponed to the end of NN search. Fig.~\ref{fig:ldg} illustrates a trail that is formed by the NN search. In the \textit{k}-nearest neighborhood of $r$, the LGD operations are applied.

Our approach is different from~\cite{dpg:wenli}, the graph diversification is undertaken in a lazy way in the sense no exhaustive comparison is involved within the \textit{k}-NN list. This scheme is therefore called as \textit{lazy graph diversification} (LGD). The three rules used to calculate occlusion factor are called as \textit{LGD rules}. Our approach is also different from the way proposed in~\cite{cvpr16:ben,pami18:yury}, in which the occluded samples ($\lambda > 0$) are simply omitted. This is infeasible in our case as it deviates the goal of \textit{k}-NN graph construction. Alternatively, the occlusion factor $\lambda$ works as an indicator of the degree of occlusion. If one sample's $\lambda$ is above the average level $\overline{\lambda}$, it is viewed as being occluded. Such kind of samples will not be visited during the fast NN search, which will be elaborated in Section~\ref{sec:nns}. Function \textit{ApplyLGD($\cdot$)} in Alg.~\ref{alg:knnbuild2}, \textit{Line 25} is responsible to fulfill \textit{LGD rules} as one sample is inserted.

\subsection{Sample Removal from \textit{k}-NN Graph}
In practice, we should allow samples to be dropped out from the \textit{k}-NN graph. A good use case is to maintain a \textit{k}-NN graph for product photos for an e-shopping website, where old-fashioned products should be withdrawn from sale. The removal of samples dynamically from the \textit{k}-NN graph is supported in our approach. If the graph is built by Alg.~\ref{alg:knnbuild1}, the removal operation is as easy as deleting the sample from \textit{k}-NN lists of its reverse neighbors and releasing its own \textit{k}-NN list. If the graph is built by Alg.~\ref{alg:knnbuild2}, before the sample is deleted, the occlusion factors of the samples living in the same \textit{k}-NN list have to be updated. Fortunately not all the samples in the list should be considered. According to \textit{LGD Rule 3}, only samples ranked after this sample should be considered. The update operations involves $k^2/2$ times distance computations in average. Given $k$ is a small constant, the time cost is much lower than fulfilling a query on the graph. 

Dynamic sample removal is not conveniently supported by other \textit{k}-NN graph construction approaches~\cite{infosys13:yury,infosys13:yury,arxiv:debatty16} or other graph-based NN search indexing structure~\cite{pami18:yury, dpg:wenli}. In HNSW~\cite{pami18:yury}, the sample removal operation may lead to the collapse of indexing structure. While this issue is not even considered in~\cite{dpg:wenli}.

\subsection{Complexity and Optimality Analysis}
For both OLGraph (Alg.~\ref{alg:knnbuild1}) and OLGraph$^+$ (Alg.~\ref{alg:knnbuild2}), the adjacency list structure is required to support the search and dynamic update. Memory for IDs, occlusion factors and distances of \textit{k} neighbors and reverse neighbors must be allocated. As a result, the upper bound of memory consumption of index is around $20{\cdot}k{\cdot}n$ bytes\footnote{This is estimated with 64bits machine.}, where \textit{n} is the scale of the dataset. Since the structure is the union of neighbors and reverse neighbors, the real memory consumption will be much lower than this bound. In terms of time complexity, both OLGraph (Alg.~\ref{alg:knnbuild1}) and OLGraph$^+$ (Alg.~\ref{alg:knnbuild2}) are essentially variants of hill-climbing. So similar as~\cite{icai11:kiana,weidong,dpg:wenli}, they are highly efficient even on high dimensional data when their intrinsic data dimension is low. However, when the intrinsic dimension is as high as several hundreds, they are only sightly better than brute-force search. Fortunately, according to our observation, the intrinsic dimension is far below that level for most of the real world data. This is particularly true for popular visual deep features.

In both OLGraph (Alg.~\ref{alg:knnbuild1}) and OLGraph$^+$ (Alg.~\ref{alg:knnbuild2}), the construction starts from a small-scale \textit{k}-NN graph of \textit{100\%} quality. The search process appends a \textit{k}-NN list of a new sample to the graph each time. At the same time, the \textit{k}-NN lists of the already inserted samples will be possibly updated when the new sample happen to be in their neighborhoods. It is therefore a win-win situation for both graph construction and NN search. Effective search procedure returns high quality \textit{k}-NN list. While high quality \textit{k}-NN graph gives a good guidance for the hill-climbing process.

Besides the size of NN list $k$, there is another parameter involved in OLGraph and OLGraph$^+$. Namely, the number of seeds $p$. Usually, the size of NN list $k$ should be no less than the intrinsic data dimension $d^*$~\cite{nips05:Levina}, which is less than or equal to the data dimension $d$. The number of seeds is usually set to be no bigger than $k$. When $d$ is very high (\textit{i.e.}, several hundreds to thousands) and $d^*$ is close to $d$, the construction process could be slow when $k$ is set to be close to $d$. In such situation, a trade-off has to be made between the quality of \textit{k}-NN graph and the efficiency of the construction.

\section{Fast NN-search on the Diversified Graph}
\label{sec:nns}
Under \textit{LGD rules}, the samples in one \textit{k}-NN list and the corresponding reverse \textit{k}-NN list are considered as being occluded when their $\lambda$ is higher than $\overline{\lambda}$ of the \textit{k}-NN list, where $\overline{\lambda}$ is the average occlusion factor of the list. This basically indicates these samples in the list are too close to other samples that they are no need to be considered during the hill-climbing. The factors of all the samples in one list are updated dynamically as long as samples are inserted/removed from the list. As the new samples are joined in the list, members that are previously occluded may become ``visible''. This is the essential difference between our approach and HNSW~\cite{pami18:yury}, in which occluded samples are removed permanently once it is identified. 

Once the occlusion factor is available, the search algorithm (Alg.~\ref{alg:nswsearch}) is accordingly modified. When the query is compared to the neighbors in \textit{r}'s \textit{k}-NN list, we only consider the samples whose $\lambda$ is no greater than the average occlusion factor $\overline{\lambda}$ of this list, where $\overline{\lambda}$ is the average occlusion factor of the \textit{k}-NN list. The NN search on the diversified \textit{k}-NN graph is in general the same as Alg.~\ref{alg:nswsearch}. In the modified NN search, a conditional judgment statement (like \textit{Line 18} in Alg.~\ref{alg:nswsearch}) is added to check whether the occlusion factor $\lambda$ of a sample is above $\overline{\lambda}$. Only the ``visible'' samples are joined in the comparison. Speed-up is expected as nearly \textit{50\%} samples in one \textit{k}-NN list are skipped during the search.

Although NN search on LGD graph is significantly more efficient than Alg.~\ref{alg:nswsearch}, it is not suitable to be adopted in the \textit{k}-NN graph construction in Alg.~\ref{alg:knnbuild2}. The close neighbors that are considered in the comparison in Alg.~\ref{alg:nswsearch} will be ignored in this modified NN search. As a consequence, the samples should be joined in \textit{k}-NN lists of each other are simply unvisited. For this reason, \textit{NNSearch}($\cdot$) (Alg.~\ref{alg:nswsearch}) in Alg.~\ref{alg:knnbuild2} is recommended when one wants to add a sample into the graph. Alternatively, NN search with LGD check is recommended as one performs pure NN search. In the experiment section, we will show that the LGD strategy leads to considerable speed-up in the NN search. It becomes competitive in comparison to the state-of-the-art approaches.

To this end, one could imagine that there are two modes in our online \textit{k}-NN graph framework, namely the construction mode and pure NN search mode. Under the pure NN search mode, \textit{ApplyLGD($\cdot$)} is not called since no \textit{k}-NN list update is invoked during the procedure. Under the construction mode, Alg.~\ref{alg:knnbuild2} is invoked.

The fast online \textit{k}-NN graph construction and NN search have been integrated into one framework. Compared to the existing \textit{k}-NN graph construction approaches~\cite{weidong, infosys13:yury,arxiv:debatty16}, fast dynamic insertion, removal as well as NN search on a \textit{k}-NN graph are all well supported. Compared to other graph-based NN indexing structures such as HNSW~\cite{pami18:yury} and DPG~\cite{dpg:wenli}, there is no offline construction stage. As a result, the cost of dynamically maintaining the indexing structure is much lower than either HNSW~\cite{pami18:yury} or DPG~\cite{dpg:wenli}. Furthermore, compared to HNSW and DPG, a diversified \textit{k}-NN graph and a \textit{k}-NN graph are maintained simultaneously\footnote{The diversified \textit{k}-NN graph is actually a subset of \textit{k}-NN graph.} in one structure. On the one hand, it guarantees the fast NN search. On the other hand, it allows the user to browse over similar contents via the links between \textit{k}-nearest neighbors.

\section{Experiments}
\label{sec:exp}
In this section, the performance of the proposed algorithms is studied both as an approximate \textit{k}-NN graph construction and a nearest neighbor search approach. In the evaluation, the performance is reported on popular evaluation datasets. The brief information about the datasets are summarized in Tab~\ref{tab:data}.

On the approximate \textit{k}-NN graph construction task, existing approaches NN-Descent~\cite{weidong} and NSW~\cite{infosys13:yury} are considered in the performance comparison, both of which are feasible for various distance metrics. On the nearest neighbor search task, the performance of the proposed search approach is studied in comparison to the representative approaches of different categories. Namely they are graph-based approaches such as NN-Descent~\cite{weidong}, HNSW~\cite{pami18:yury}. The representative locality sensitive hash approach SRS~\cite{srs14} is considered. For quantization based approach, product quantizer (PQ)~\cite{JDS11} is incorporated in the comparison. FLANN~\cite{pami14:flann} and Annoy~\cite{annoy} are selected as the representative tree partitioning approaches. Both of them are popular NN search libraries in the literature.

\subsection{Evaluation Protocol}
Eight real world datasets are adopted to evaluate the performance of both \textit{k}-NN graph construction and nearest neighbor search. These datasets are derived form real world images, text data or itemset. The top-\textit{1} (\textit{recall@1}) and top-\textit{10} (\textit{recall@10}) recalls on each dataset are studied under different metrics such as $\textit{l}_2$, \textit{Cosine}, \textit{Jaccard} and $\kappa^2$. Given function $R(i,k)$ returns the number of truth-positive neighbors at top-\textit{k} NN list of sample $i$, the recall at top-$k$ on the whole set is given as
\begin{equation}
	recall@k=\frac{\sum_{i=1}^n{R(i,k)}}{n{\times}k}.
\label{eval:recall}
\end{equation}

\begin{table}[t]
\caption{Summary on Datasets used for Evaluation}
\scriptsize{
\begin{center}
	\begin{tabular}{|l||c|c|c|c|c|} \hline
	Name & 	$n$ & $d$ & \# Qry & m($\cdot$,$\cdot$) & Type \\ \hline\hline
	SIFT1M~\cite{JDS11} & $1{\times}10^6$ & $128$ & $1{\times}10^4$ & $\textit{l}_2$ & SIFT~\cite{ijcv04:sift} \\ \hline
	SIFT10M~\cite{JDS11} & $1{\times}10^7$ & $128$ & $1{\times}10^4$ & $\textit{l}_2$ & SIFT \\ \hline
	GIST1M~\cite{civr09:jegou} & $1{\times}10^6$ & $960$ & $1{\times}10^3$ & $\textit{l}_2$ & GIST~\cite{civr09:jegou} \\ \hline
	GloVe1M~\cite{glove14} & $1{\times}10^6$ & $100$ & $1{\times}10^3$ & \textit{Cosine} & Text \\ \hline
	NUSW~\cite{nusw09} & $22,660$ & $500$ & $1{\times}10^3$ & $\textit{l}_2$ & BoVW~\cite{iccv03:sivic} \\ \hline
	NUSW~\cite{nusw09} & $22,660$ & $500$ & $1{\times}10^3$ & ${\kappa}^2$ & BoVW \\ \hline
	YFCC1M~\cite{yfcc100m} & $1{\times}10^6$ & $128$ & $1{\times}10^4$ & $\textit{l}_2$ & Deep Feat. \\ \hline
	Kosarak~\cite{aumuller2017ann} & $7.4{\times}10^4$ & 27983 & $1{\times}10^3$ & \textit{Jaccard} & Itemset \\ \hline
	\end{tabular}
\end{center}
}
\label{tab:data}
\end{table}

Besides \textit{k}-NN graph quality, the construction cost is also studied by measuring the scanning rate~\cite{weidong} and time cost of each approach. Given $C$ is the total number of distance computations in the construction, the scanning rate is defined as
\begin{equation}
c=\frac{C}{n{\times}(n-1)/2}.
\label{eval:scan}
\end{equation}

For each dataset, another \textit{1,000} or \textit{10,000} queries of the same data type are prepared. The NN search quality is measured by the top-\textit{1} recall for the first nearest neighbor. The search quality is reported along with the number of queries per second. All the codes of different approaches considered in this study are compiled by g++ \textit{5.4}. In order to make our study to be fair, we disable all the multi-threads, SIMD and pre-fetching instructions in the codes since not all the original codes are optimized with these techniques. All the experiments are executed on a PC with \textit{3.6}GHz CPU and \textit{32}G memory setup.

\begin{table*}[htb]
	\caption{Scanning rates and time consumption of NN-Descent, NSW, OLGraph and OLGraph$^+$ on eight datasets}
	\label{tab:scall}
	\begin{center}
		\begin{tabular}{|c| |c|r| |c|r| |c|r| |c|r|} \hline
		\multirow{2}{*}{Dataset} &
		\multicolumn{2}{c||}{NN-Descent} & \multicolumn{2}{c||}{NSW} & \multicolumn{2}{c||}{OLGraph} & \multicolumn{2}{c|}{OLGraph$^+$} \cr\cline{2-9}
		&Scanning rate & Time(s) & Scanning rate & Time(s) & Scanning rate & Time(s) & Scanning rate & Time(s)\cr
		\hline
		SIFT1M & 0.01856 & 1278 & 0.00899 & 999 & 0.00699 & 1015 & \textbf{0.00606} & \textbf{997} \cr \hline
		SIFT10M & 0.00220 & 15672 & 0.00130 & 16639 & 0.00090 & \textbf{15286} & \textbf{0.00086} & 15667 \cr \hline
		GIST1M & 0.02406 & 11497 & 0.02029 & 10140 & 0.02140 & 12186 & \textbf{0.01421} & \textbf{8412} \cr \hline
		YFCC1M & 0.01674 & 1206 & 0.01143 & 1288 & \textbf{0.00732} & \textbf{1077} & 0.00765 & 1247 \cr \hline
		GLoVe1M & 0.01881 & \textbf{1463} & 0.01453 & 2010 & 0.01124 & 1920 & \textbf{0.01101} & 2011 \cr \hline
		NUSW-$l_2$ & 0.07655 & 1377 & 0.07624 & 1517 & 0.08749 & 2010 & \textbf{0.05149} & \textbf{1207} \cr \hline
		NUSW-$k^2$ & 0.07934 & 7546 & 0.08490 & 6592 & 0.08970 & 9441 & \textbf{0.05948} & \textbf{6193} \cr \hline
		Kosarak & 0.21321 & 358 & 0.12524 & 317 & 0.12202 & 351 & \textbf{0.11200} & \textbf{302} \cr \hline
		\end{tabular}
	\end{center}
\end{table*}

\subsection{Approximate \textit{k}-NN Graph Construction}
\begin{figure}[htb]
	\begin{center}
	\subfigure[Recall@1]
    {\includegraphics[width=0.483\linewidth]{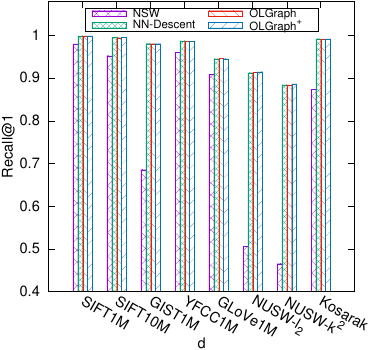}}
    \hspace{0.02in}
	\subfigure[Recall@10]
	{\includegraphics[width=0.483\linewidth]{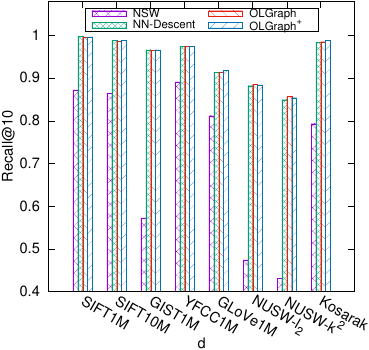}}	
	\caption{Top-\textit{1} and Top-\textit{10} recall of \textit{k}-NN graphs produced by NSW, NN-Descent, OLGraph and OLGraph$^+$ on eight datasets.}
	\label{fig:krecall}
\end{center}
\end{figure}

In the first experiment, the performance of approximate \textit{k}-NN graph construction is studied when the $k$-NN search algorithm is employed as a graph construction approach. Eight datasets derived from real world data are adopted. Among them, NUSW is tested under both $l_2$ and $\kappa^2$ distance measures, and Kosarak uses $Jaccard$ distance measure. In the evaluation, the performance of OLGraph (Alg.~\ref{alg:knnbuild1}) and OLGraph$^+$ (Alg.~\ref{alg:knnbuild2}) is compared to NN-Descent~\cite{weidong} and NSW~\cite{infosys13:yury}. NN-Descent is recognized as the most effective approximate \textit{k}-NN graph construction approach that works in the generic metric spaces, and NSW is an online approach to construct a navigable small world graph for \textit{k}-nearest neighbor search. The \textit{k}-NN graph can be derived from NSW graph by only keeping the first \textit{k} nearest neighbors. In the test, the parameter \textit{k} in NN-Descent is fixed to \textit{40} for all the datasets, and the parameters of OLGraph and OLGraph$^+$ are tuned to reach similar recalls as NN-Descent. Since it is difficult for NSW to reach high recall \textit{k}-NN graph on most of the datasets, its performance is reported on the level that its time cost is similar as other approaches. The scanning rates and time consumption of all four approaches are reported in Tab.~\ref{tab:scall}. While the top-\textit{1} and top-\textit{10} recalls of all the approaches are shown in Fig.~\ref{fig:krecall}.

As seen from the table, in the most of cases, the scanning rates from OLGraph and OLGraph$^+$ are much lower than that of NN-Descent as their graph quality is maintained on the similar level. This basically indicates much less distance computations are involved with OLGraph and OLGraph$^+$. Compared to NN-Descent, OLGraph and OLGraph$^+$ avoid repetitive distance computations between any sample pairs. As the distance computation is the most computationally intensive operation in all approaches, OLGraph and OLGraph$^+$ are expected to be much faster. However, the CPU cache structure is more friendly to NN-Descent since its distance computations are taken place in a local at each moment. The computation costs from OLGraph are not considerably lower than NN-Descent as is shown in Tab.~\ref{tab:scall}. Compared to OLGraph, OLGraph$^+$ shows considerably lower time costs than NN-Descent due to its much lower scanning rate. In particular, it has a greater advantage on high-dimensional datasets like GIST1M and NUSW. 

As shown in Fig.~\ref{fig:krecall}, \textit{k}-NN graphs produced by NSW show considerably poorer quality than the other approaches when they take similar time costs. Although NSW and OLGraph are conceptually similar to each other, they are essentially different from each other. In both OLGraph and OLGraph$^+$, the new query is attempted to insert into the neighborhoods of all the visited samples. In NSW, only the neighborhoods of returned top-\textit{k} samples to the query are updated when a new query is joined in. As a result, a sample outside of top-\textit{k} ranking has no chance to add the new query into its neighborhood. For this reason, the graph quality is low. This in turn degrades its performance as an NN search indexing structure.

\subsection{Nearest Neighbor Search}
In our second experiment, the NN search performance is compared to two representative graph-based approaches NN-Descent~\cite{weidong} and HNSW~\cite{pami18:yury}, which work in generic metric spaces. All the approaches use the similar hill-climbing search procedure. They are different from each other mainly in the graphs upon which the search procedure is undertaken. For convenience, the NN search approaches based on the graph constructed by OLGraph and OLGraph$^+$ are given as OLGraph and OLGraph$^+$ respectively. The NN search on OLGraph graph is based on Alg.~\ref{alg:nswsearch}, while NN search on OLGraph$^+$ graph is the approach presented in Section~\ref{sec:nns}, in which the LGD check is adopted. In the experiment, parameter \textit{k} in NN-Descent is fixed to \textit{40} for all the datasets to be in line with the experiments in ~\cite{dpg:wenli}. OLGraph searches over graph which is a merge of \textit{k}-NN graph and its reverse \textit{k}-NN graph that are produced by Alg.~\ref{alg:knnbuild1}. NN search in OLGraph$^+$ is on a \textit{k}-NN graph merged with its reverse \textit{k}-NN graph that have been diversified online by \textit{LGD rules} (Alg.~\ref{alg:knnbuild2}). The parameter \textit{k} in OLGraph and OLGraph$^+$ is also fixed to \textit{40} for all the datasets. While for HNSW, the search is undertaken on a hierarchical small world graph. The graph maintains the links between the close neighbors as well as the long range links to the remote neighbors that are kept in a hierarchy. The parameter $M$ in HNSW is fixed to \textit{20}. The edges kept for each sample in the bottom-layer is \textit{40}. Its size of NN list is therefore on the same level as NN-Descent, OLGraph and OLGraph$^+$.

\begin{figure*}[htb]
	\begin{center}
	\subfigure[SIFT1M]
    {\includegraphics[width=0.2\linewidth]{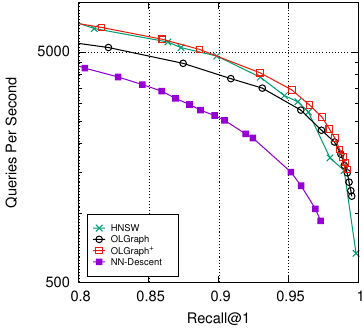}}
    \hspace{0.06in}
	\subfigure[SIFT10M]
    {\includegraphics[width=0.2\linewidth]{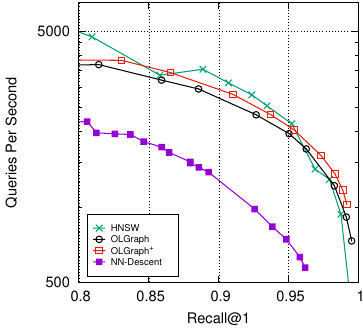}}
    \hspace{0.06in}
	\subfigure[GIST1M]
	{\includegraphics[width=0.2\linewidth]{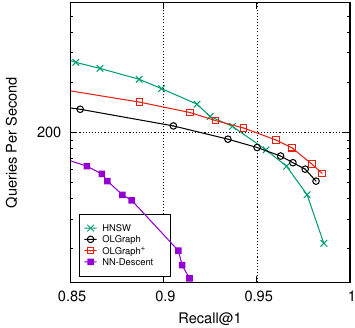}}
    \hspace{0.06in}
	\subfigure[YFCC1M]
	{\includegraphics[width=0.2\linewidth]{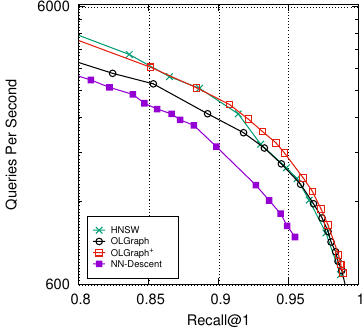}}\\
	\subfigure[NUSW-$l_2$]
	{\includegraphics[width=0.2\linewidth]{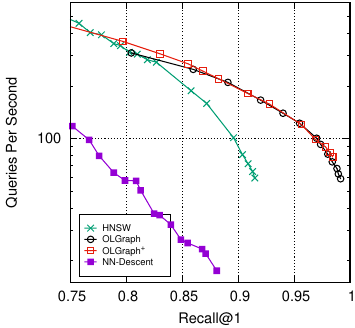}}
    \hspace{0.06in}
	\subfigure[NUSW-$\kappa^2$]
	{\includegraphics[width=0.2\linewidth]{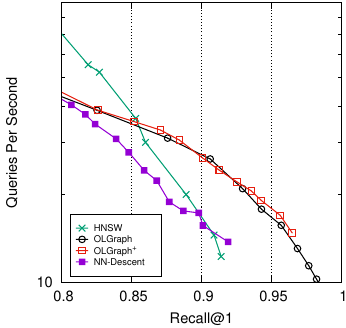}}
    \hspace{0.06in}
	\subfigure[GloVe1M]
	{\includegraphics[width=0.2\linewidth]{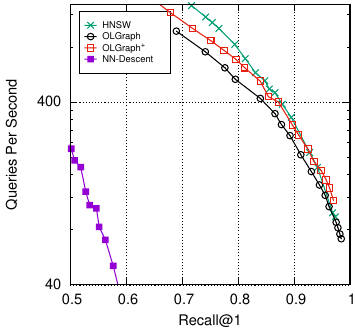}}
    \hspace{0.06in}
	\subfigure[Kosarak]
	{\includegraphics[width=0.2\linewidth]{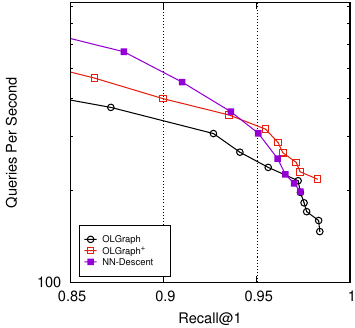}}
	\caption{The NN search performance on eight datasets. Four graph-based approaches are considered in this study. OLGraph and OLGraph$^+$ are the approaches proposed in this paper. Because the source code of HNSW does not support sparse matrices, it does not participate in the comparison on Kosarak dataset.}
	\label{fig:sprecall}
\end{center}
\end{figure*}

The search performance on eight datasets is shown in Fig.~\ref{fig:sprecall}. Among all the graph-based approaches, the relative better performance is observed from HNSW, OLGraph and OLGraph$^+$ over NN-Descent. The performance boost mainly owes to the use of reserve \textit{k}-NN list and the introduction of graph diversification. The trend of OLGraph performance curve is similar as that of OLGraph$^+$, whereas OLGraph$^+$ outperforms OLGraph on most of the datasets. The performance superiority of OLGraph$^+$ largely owes to the LGD strategy. Compared to HNSW, OLGraph$^+$ shows increasingly better performance as the recall is set to high level. It is mainly because OLGraph$^+$ performs graph diversification on a high quality approximate \textit{k}-NN graph. In contrast, limited by the search framework, HNSW performs the diversification on a dynamically diversified graph.

Compared the result presented in Fig.~\ref{fig:sprecall}(a) to the one presented in Fig.~\ref{fig:sprecall}(b), the high scalability is observed on SIFT data by the proposed approach. As seen in the figure, the size of reference set has been increased by one magnitude, while the time cost only increases from \textit{0.21}ms (per query) to \textit{0.32}ms (per query), when the search quality is maintained on \textit{0.9} level. Similar high scalability is also observed on deep features \textit{i.e.}, YFCC1M (Fig.~\ref{fig:sprecall}(d)). This is good news given the deep features have been widely adopted in various applications nowadays. In contrast, such kind of high speed-up is not achievable on NUSW, GloVe1M and GIST1M. It is clear to see that the efficiency of graph-based approaches is partly related to the intrinsic data dimension~\cite{weidong,cvpr16:ben}. When the intrinsic data dimension is low, with the guidance of a \textit{k}-NN graph or a relative \textit{k}-NN graph, the hill-climbing search is actually undertaken on the subspaces where most of the data samples are embedded. Due to the low dimensionality of these subspaces, the search complexity is lower than it seemingly is. This is one of the major reasons that the graph-based approaches exhibit superior performance over other type of approaches.

\subsection{Comparison to state-of-the-art \textit{k}-NN Search}
\begin{figure}[htb]
	\begin{center}
	\subfigure[Recall@1=0.8]
    {\includegraphics[width=0.48\linewidth]{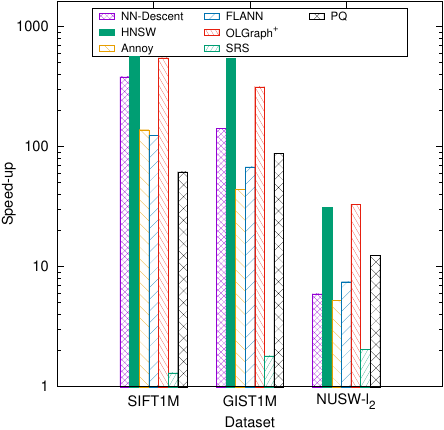}}
	\subfigure[Recall@1=0.9]
    {\includegraphics[width=0.47\linewidth]{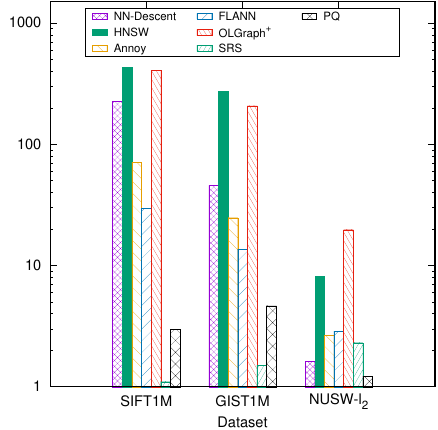}}
	\caption{The NN search performance on three datasets ranging from ``easy'' to ``hard'' (best viewed in color). Seven representative approaches in the literature are considered in the comparison. Figures (a) and (b) report the speed-up that one approach could achieve when the top-\textit{1} recall is on \textit{0.8} and \textit{0.9} levels respectively.}
	\label{fig:methods}
\end{center}
\end{figure}

Fig.~\ref{fig:methods} further compares our approach with the most representative approaches of different categories in the literature. Besides aforementioned HNSW and NN-Descent, approaches considered in the comparison include tree partitioning approaches Annoy~\cite{annoy} and FLANN~\cite{pami14:flann}, locality sensitive hashing approach SRS~\cite{srs14}, and vector quantization approach product quantizer (PQ)~\cite{JDS11}. In the figures, the speed-up relative to exhaustive search that each approach achieves is reported when \textit{recall@1} is set to \textit{0.8} and \textit{0.9} levels. For PQ, it is impossible to achieve top-\textit{1} recall above \textit{0.5} due to its heavy quantization loss. As an exception, its recall is measured at top-\textit{16} for SIFT1M and NUSW, and measured at top-\textit{128} for GIST1M.

As shown in the figure, the best results come from graph-based approaches. This observation is consistent across different datasets. The speed-up of all the approaches drops as the \textit{recall@1} rises from \textit{0.8} to \textit{0.9}. The speed-up degradation is more significant for approaches such as PQ and FLANN. No considerable speed-up is observed for SRS on any of the datasets. This basically indicates SRS is not suitable for the tasks which require high NN search quality. Another interesting observation is that the performance gap between graph-based approaches and the rest is wider on the ``easy'' dataset than that of ``hard''. Compared to the approaches of other categories, the NN search based on the graph takes advantage of the latent subspace structures in a dataset. Since the intrinsic dimension of ``easy'' dataset is low~\cite{cvpr16:ben}, the hill-climbing is actually undertaken on these low-dimensional subspaces. The higher is the ratio between data dimension and intrinsic dimension, the higher is the speed-up that graph-based approaches achieve. In contrast, there is no specific strategy in other type of approaches capitalizes on such latent structures in the data.

On the one hand, the high search speed-up is observed from OLGraph$^+$ on data types such SIFT, GIST and deep features. With such efficiency, it is possible to realize a search system with instant response on \textit{100} million level dataset by a single PC. On the other hand, it is still too early to say the problem of NN search on high-dimensional data has been solved. As shown on NUSW dataset, where both the data dimension and intrinsic data dimension are high, the efficiency achieved from all the approaches is still limited. As pointed out in our another work~\cite{arxiv:pclin19}, the difficulty faced in this case is directly linked to ``curse of dimensionality'', which will remain as an open issue.

\section{Conclusion}
\label{sec:end}
We have presented our solution for both approximate \textit{k}-NN graph construction and nearest neighbor search. These two issues have been addressed under a unified framework. Namely, the NN search and NN graph construction are designed as an interdependent procedure that one is built upon another. The advantages of this design are several folds. First of all, the approximate \textit{k}-NN graph construction is treated as an online procedure. It allows the samples to be inserted in or dropped out from the graph dynamically, which is not possible from most of the existing solutions. Moreover, no sophisticated indexing structure is required to support this online approach. Furthermore, the solution has no specification on the distance measure, which makes it a generic approach both for \textit{k}-NN graph construction and NN search. Superior performance is observed on both \textit{k}-NN graph construction and nearest neighbor search tasks under various test configurations.

\ifCLASSOPTIONcompsoc
  \section*{Acknowledgments}
\else
  \section*{Acknowledgment}
\fi

This work is supported by National Natural Science Foundation of China under grants 61572408 and 61972326, and the grants of Xiamen University 20720180074.

\bibliographystyle{ieeetr}
\bibliography{wlzhao}

\end{document}